\documentclass[aps,prb,twocolumn,superscriptaddress,showpacs,floatfix]{revtex4}

\usepackage{multirow}
\usepackage{graphicx}
\usepackage{dcolumn}
\usepackage{bm}
\usepackage{stmaryrd}
\usepackage{latexsym}
\usepackage{amssymb}
\usepackage{amsfonts}
\usepackage{amsmath}

\begin{document}

\title{Electronic and Magnetic Structures of Chain Structured Iron Selenide Compounds}

\author{Wei Li}
\affiliation{Department of Physics, Fudan University, Shanghai 200433, China}

\author{Chandan Setty}
\affiliation{Department of Physics, Purdue University, West Lafayette, Indiana 47907, USA}

\author{X. H. Chen}
\affiliation{Hefei National Laboratory for Physical Science at Microscale and Department of Physics,
University of Science and Technology of China, Hefei, Anhui 230026, China}

\author{Jiangping Hu}
\email{jphu@iphy.ac.cn}
\affiliation{Beijing National Laboratory for Condensed Matter Physics, and Institute of Physics, Chinese Academy of Sciences, Beijing 100190, China}
\affiliation{Department of Physics, Purdue University, West Lafayette, Indiana 47907, USA}

\date{\today}


\begin{abstract}
Electronic and magnetic structures of iron selenide compounds Ce$_2$O$_2$FeSe$_2$ ($2212^*$) and BaFe$_2$Se$_3$($123^*$) are studied by the first-principles calculations. We find that while all these compounds are composed of one-dimensional (1D) Fe chain (or ladder) structures, their electronic structures are not  close to be quasi-1D. The magnetic exchange couplings between two nearest-neighbor (NN) chains in $2212^*$ and between two NN two-leg-ladders in $123^*$ are both antiferromagnetic (AFM), which is consistent with the presence of significant third NN AFM coupling, a common feature  shared in other iron-chalcogenides,  FeTe ($11^*$) and K$_y$Fe$_{2-x}$Se$_2$ $(122^*)$. In  magnetic ground states, each Fe chain of $2212^*$ is ferromagnetic and each two-leg ladder of $123^*$ form a block-AFM structure. We suggest that all magnetic structures in iron-selenide compounds can be unified into an extended $J_1$-$J_2$-$J_3$ model. Spin-wave excitations of the model are calculated and can be tested by future experiments on these two systems.
\end{abstract}

\maketitle

\section{introduction}

The newly discovered $122^*$, $A_y$Fe$_{2-x}$Se$_2$, iron-chalcogenide superconductors\cite{Jiangang_Guo,Hechang_Lei,Maziopa,RHLiu,MinghuFang} have attracted enormously interests. Like iron-pnictide high temperature superconductors, the FeSe-based superconductors have the same robust tetrahedral layers structure.  However, there are a number of distinct intriguing physical properties which are noticeably absent in iron-pnictide materials. Such as antiferromagnetically (AFM) ordered insulating phases\cite{MinghuFang,ZGChen}, extremely high N\'{e}el transition temperatures\cite{WBao,VYu}, and the presence of intrinsic Fe vacancies and ordering\cite{ZWang,Zavalij,XunWang_Yan1,Chaocao}. In addition, the Fermi surface (FS) topologies of superconducting compounds are very different from previously known superconducting Fe-pnictides. Both band structures calculations\cite{Nebrasov,XunWang_Yan,Shein,ChaoCao} and angle resolved photoemission spectroscopy studies\cite{YZhang,TQian} indicated that only the electron pockets are present in the superconducting compounds, while the hole pockets around $\Gamma$ point observed in iron-pncitide counterparts sink below the Fermi level, indicating that the inter-pocket scattering between the hole and electron pockets is not an essential ingredient for superconductivity.

\begin{figure}
\includegraphics[width=8cm]{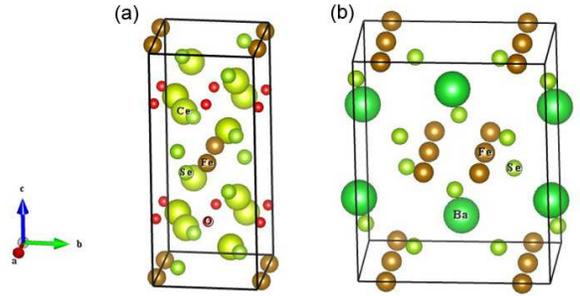}
\caption{(Color online) Calculated crystal structures of the iron selenide compounds: (a) $2212^*$, which consists of one chain of FeSe$_4$ tetrahedra structure, and (b) $123^*$, which consists of two-leg-ladders of edge-sharing FeSe$_4$ tetrahedra structure, respectively.}\label{fig:fig1}
\end{figure}

Following the  discovery of the $122^*$ iron-chalcogenide, two new materials, Ce$_2$O$_2$FeSe$_2$ ($2212^*$)\cite{McCabe} and BaFe$_2$Se$_3$($123^*$)\cite{Maziopa2011,Caron,saparov,JMCaron}, have been synthesized. In the $2212^*$, each Fe-layer is composed of coupled one-dimensional (1D) Fe-chains and  in the $123^*$, it is structured by coupled two-leg ladders. An analogy of these materials can be made to those in cuprates, such as a spin-ladder system Sr$_{14-x}$Ca$_{x}$Cu$_{24}$O$_{41}$\cite{McCarron1988,Siegrist1988} (x=11.5-15.5) and a double chain system Pr$_2$Ba$_4$Cu$_7$O$_{15-\delta}$\cite{Nakano}. $123^*$\cite{Maziopa2011,Caron,saparov,JMCaron}, which has been investigated intensively. It was also reported that the $123^*$ may be superconducting\cite{Maziopa2011}.

In this paper, we present the theoretical study of the electronic band structures and magnetic orders in these iron selenide systems featured with low-dimensional iron structures. We investigate two materials including $2212^*$ and $123^*$ and show that while all these compounds are composed of 1D Fe chain structures, their electronic structures are not close to be quasi-1D. Their FS still exhibit two dimensional or even three dimensional topologies. We calculate their magnetic ordered ground states. In $2212^*$, the magnetic order ground state is a collinear-AFM (CAF), similar to iron-pncitides. In $123^*$, the magnetic structure is a block-AFM (BAF), similar to KFe$_2$Se$_2$\cite{Liwei}. The magnetic exchange couplings between two nearest neighbor (NN) chains in $2212^*$ and between two NN two-leg-ladders in $123^*$ are both AFM, which is consistent with the presence of significant third NN AFM coupling, $J_3$, in FeTe ($11^*$) and $122^*$. This result suggests that  all magnetic structures in iron selenide compounds can be unified into an extended $J_1$-$J_2$-$J_3$ model. We also calculate spin-wave excitations of the model which can be tested in future experiments on these two systems.

\section{Theories and Results}

\subsection{First-Principles Calculations}

\begin{table}
\caption{Structural parameters, density of states at the Fermi level $N(E_F)$ (in the (eV)$^{-1}$ units per Fe atom) and the calculated specific heat coefficient $\gamma_0[mJ/(K^2mol)]$, and Pauli susceptibility $\chi_0(10^{-9}m^3/mol)$ for iron selenide compounds in NM state. The lattice parameters and the internal coordinates are all optimized within energy minimization.}
\label{tableI}
\begin{tabular}{lcccccc}
\hline
\hline
  & ~~~$a$(\AA)~~~ & ~~$b$(\AA)~~ & ~~$c$(\AA)~~ &  $N(E_F)$ & $\gamma_0$ & $\chi_0$~~  \\
\hline
~~$2212^*$~~  & 5.5508 & 5.6794 & 16.2566 & 6.7293 & 31.7196& 5.4546  \\
~~$123^*$~~       & 5.3821 & 9.1123 & 11.2096 & 0.7650 & 3.6060 & 0.6201  \\
\hline
\hline
\end{tabular}
\end{table}

We perform the first-principles calculations on the iron selenide compounds: $2212^*$, which has 1D chains of edge-shared FeSe$_4$ tetrahedra structure and $123^*$, which consists double chains (two-legged ladders) of edge-shared FeSe$_4$ tetrahedra structure. The crystal structures are shown in Fig. \ref{fig:fig1}. In our calculations the projected augmented wave method\cite{PBE} as implemented in the VASP code\cite{VASP}, and the Perdew-Burke-Ernzerhof exchange correlation potential\cite{ECP} was used. All atomic positions and the lattice constants are allowed to relax simultaneously to minimize the energy only for nonmagnetic (NM) state. The experimental crystal structures\cite{McCabe,Caron} are used for calculating magnetic states. A 500eV cutoff in the plane wave expansion ensures the calculations converge to $10^{-5}$ eV, and all atomic positions and the lattice constants were optimized until the largest force on any atom was 0.005eV/\AA. To properly describe the strong electron correlation in the $4f$ rare earth element Ce, the LDA plus on-site repulsion $U$ method (LDA+$U$) was employed with the effective U value ($U_{eff}$ = $U - J$) of 12.0eV for the compound $2212^*$, where the $U_{eff}$ value has been reported in the previous work\cite{Pourovskii} of CeOFeAs. The results are also checked for consistency with varying $U_{eff}$ values. We do not apply $U_{eff}$ to the itinerant Fe-$3d$ states.

\begin{figure}[tbp]
\includegraphics[width=8.5cm]{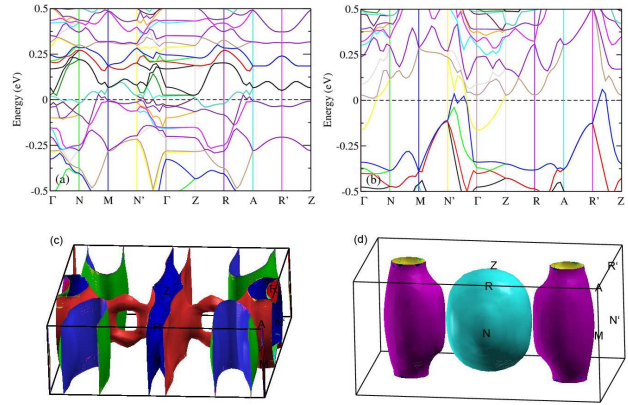}
\caption{(Color online) Electronic structures of the iron selenide compounds in the NM state: The band structure of $2212^*$ (a), $123^*$ (b), and the corresponding FS (c) and (d), respectively. The Fermi energy is set to zero.}\label{fig:fig2}
\end{figure}

First, we focus on the electronic structures of the iron selenide compounds and their dependence on the structural factors. For this purpose, full structural optimization of the these compounds were performed both over the lattice parameters and the atomic positions including the internal coordinate $z$ of Se atom by the energy minimization. All these results (not included the internal coordinate) are summarized in the Table \ref{tableI}. Actually, both iron selenide compounds of the lattice parameters optimized in our NM calculations are found smaller by about 2\% than the ones in experimental values\cite{McCabe,Maziopa2011,Caron,saparov}. In addition, the density of states $N(E_F)$ at the Fermi energy are also calculated, and the corresponding electronic specific heat coefficient $\gamma_0$ and Pauli susceptibility $\chi_0$ are all listed in Table \ref{tableI}.

Figure \ref{fig:fig2} shows the NM state band structure and FS of both iron selenide compounds. As we can see that there are three bands crossing the Fermi level for both iron selenide compounds. Although both iron selenide compounds are composed of 1D Fe chain (ladder) structures and exhibit quasi-1D characters, their FSs still exhibit two-dimensional or even three dimensional complex topologies. Their NM state electronic structures are very distinct from that of the iron selenide superconductor KFe$_2$Se$_2$\cite{Shein,XunWang_Yan,ChaoCao,YZhang}. Therefore, if superconductivity exists in these compounds, it provides a new  playground to test theoretical mechanisms.

\begin{figure}[tbp]
\includegraphics[width=8cm, height=7.5cm]{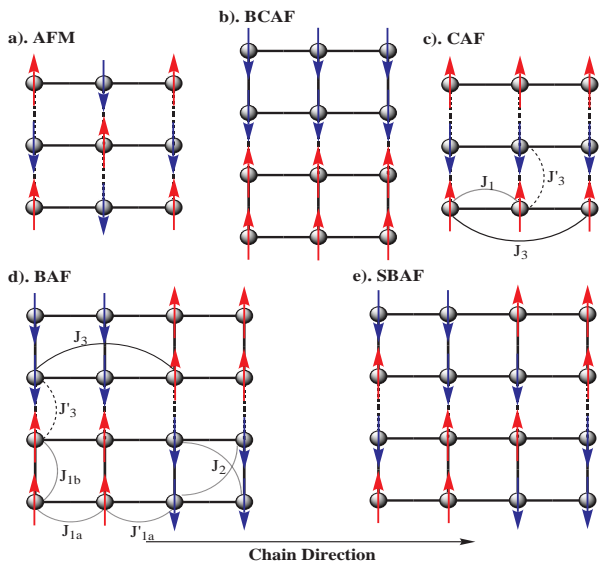}
\caption{(Color online) Schematic top view of five possible magnetic orders in the Fe-Fe square layer of the iron selenide compounds: (a) AFM N\'{e}el order in which the nearest neighboring Fe moments are AFM ordered; (b) Bicollinear-AFM (BCAF) order (the chain direction is changed into vertical direction for $2212^*$); (c) CAF order in which the Fe moments are FM ordered along the chain direction and AFM ordered across the chains direction; (d) BAF consisting of FM Fe$_4$ plaquettes tiled AFM along the chain direction; (e) Staggered-BAF (SBAF) configuration with FM diagonal double stripes that are also tiled AFM.}\label{fig:fig3}
\end{figure}

\begin{figure}[tbp]
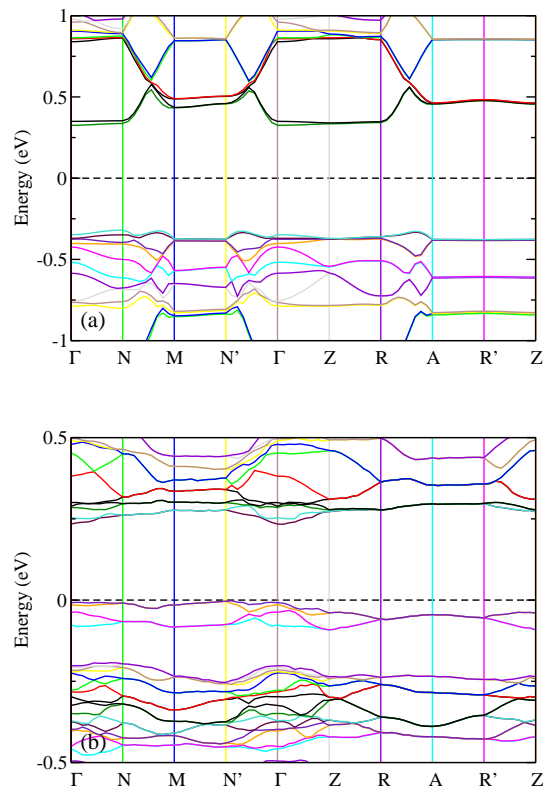

\begin{center}
\tabcolsep=0cm
\begin{tabular}{cc}
\includegraphics[bb=0 35 730 580 width=8cm, height=5.5cm]{Fig4a.eps} \\
\includegraphics[bb=0 35 730 580 width=8cm, height=5.5cm]{Fig4b.eps}
\end{tabular}%
\end{center}
\caption{(Color online) (a) Electronic band structure of the CAF state in $2212^*$ with an energy band gap $E_g$ = 0.64eV; (b) Electronic band structure of the BAF state in $123^*$ with an energy band gap $E_g$=0.24eV. The Fermi energy is set to zero.}
\label{fig:fig4}
\end{figure}

Because the NM state is strongly unstable against moment formation, we turn to study the magnetic structures in both iron selenide compounds. The six different possible magnetic configurations, as shown in Fig. \ref{fig:fig3} [ferromagnetic (FM) state has not been included] are all calculated. In Table \ref{tableII}, we list the energies of different magnetic states. For $2212^*$, it is shown that the CAF is the lowest energy state. In the CAF state of $2212^*$, spins are FM for each Fe-chains and are AFM between two NN chains. The calculated magnetic moment around each Fe ion is found to be about 3.12$\mu_B$, which is well consistence with experimental results 3.33$\mu_B$ at low temperature 12K\cite{McCabe}. Furthermore, the calculated band structure of CAF state is shown that $2212^*$ is a semiconductor with an energy band gap of around 0.64eV, as shown in Fig. \ref{fig:fig4}(a), which is also well consistence with reported experimental results\cite{McCabe}. For $123^*$, the BAF state is the lowest energy state. In the BAF state of $123^*$, for each two-leg ladder, four spins group together become a superunit. Spins between two NN units are AFM. The coupling between two NN two-leg ladders is also AFM. The moment around each Fe is about 2.85$\mu_B$ for $123^*$, and the electronic band structure calculated shows a semiconductor with an energy band gap $E_g$=0.24eV, as shown in Fig. \ref{fig:fig4}(b). The very small energy difference between FM state and CAF state is indicative of weak AFM coupling between different chains, and the energy difference between AFM state and CAF state is indicative of strong FM coupling along the chain direction in $2212^*$. Similarly, in $123^*$ system, the small energy difference between FM state and BCAF state is also indicative of weak AFM coupling between different ladders, and the energy difference between BAF state and BCAF state indicative of four Fe atom plackets along the ladders.

\begin{table}
\caption{Energetic and magnetic properties of the $2212^*$ and $123^*$. Results in the magnetic states configurations, as shown in Fig. \ref{fig:fig3} using experimental crystal structures\cite{McCabe,Caron}. $\Delta E$ is the total energy difference per iron atom in reference to
the FM state, and $m_{Fe}$ is the local magnetic moment on Fe.}
\label{tableII}
\begin{tabular}{lcccccc}
\hline
\hline
  & ~~~$2212^*$~~~ & ~~~$123^*$~~~   \\
\hline
\hline
  & ~~~~~$\Delta E$(eV)/$m_{Fe}$($\mu_B$)~~~~~&~~~~~$\Delta E$(eV)/$m_{Fe}$($\mu_B$)~~~~~\\
\hline
FM      & 0/3.13                   & 0/2.58               \\
AFM     & 0.2184/3.11              &-0.1131/2.38          \\
CAF     &\textbf{-0.0118}/3.12     &-0.1560/2.77          \\
BCAF    &0.0944/3.13               &-0.0139/2.55          \\
BAF     &  $-$                    &\textbf{-0.1615}/2.85  \\
SBAF    &  $-$                    &-0.1514/2.74           \\
\hline
\hline
\end{tabular}
\end{table}

Nevertheless, these results are consistent with magnetic exchange couplings obtained in other iron-chalcogenides FeTe and K$_{0.8}$Fe$_{1.6}$Se$_2$, where an FM NN coupling $J_1$, an AFM next nearest neighbor (NNN) coupling $J_2$, and an third NN AFM $J_3$ are necessary in describing magnetic orders. As we will show in next subsection, the magnetic orders of both materials can be obtained within models with the similar exchange coupling parameters. The AFM couplings between two NN chains in the $2212^*$ and between two NN two-leg ladders in the $123^*$ are exactly the third NN AF coupling, $J_3$. The BAF order within each two-leg ladder can also be naturally understood from these couplings. Therefore, overall, the magnetism of all iron-chalcogenides can be unified into an effective model that includes local magnetic exchange couplings as suggested in \onlinecite{JPHu} and \onlinecite{JPHu2}. The values of exchange couplings can not be accurately determined since the result depends on the selection of the magnetic configurations\cite{Liwei}.

\subsection{Magnetic Model for Ce$_2$O$_2$FeSe$_2$}

Following  above results  and effective models derived for other iron-chalcogenides, we construct the following model to describe this material
\begin{eqnarray}
{\hat H} = J_1 \sum_{\langle i,j\rangle}\vec{S}_i\cdot\vec{S}_j+J_3 \sum_{\langle\langle i,j\rangle\rangle} \vec{S}_i\cdot \vec{S}_j + J'_3 \sum_{\langle i,j\rangle} \vec{S}_i\cdot \vec{S}_j,
\label{eq:one}
\end{eqnarray}
where $J_1$ and $J_3$ are the NN and next NN intrachain magnetic exchange couplings and $J'_3$ is the NN interchain magnetic exchange coupling, as shown in Fig. \ref{fig:fig3}(c).

The classical ground state of the Hamiltonian  can be obtained exactly. In general, the classical energy is given as (for simplicity, we take $S=1$)
\begin{eqnarray*}
E_c  =  2 J_1\cos Q_x + 4 J_3\left(\cos^2Q_x -1/2\right)+ 2 J'_3\cos Q_y,
\end{eqnarray*}
where ($Q_x$, $Q_y$) are the magnetic order wavevectors which can be viewed as the relative polarization angles between two NN intrachain spins  and  interchain spins respectively.
The CAF phase is obtained when $J_3<|J_1|/4$ and $J'_3>0$.

In this state, we perform a linear spin wave analysis for this material in the classical limit. To do this we use the usual linearized Holstein-Primakoff transformation from spin operators to magnon operators which read as

\begin{eqnarray*}
S_i^x = \sqrt{\frac{S}{2}}(b_i + b_i^{\dagger}); S_i^y = -i \sqrt{\frac{S}{2}}(b_i - b_i^{\dagger}); S_i^z = S - b_i^{\dagger}b_i
\end{eqnarray*}
where $i$ runs over all the lattice sites. Performing a fourier transform, the spin wave excitations of the model is given by
\begin{equation}
{\hat H} = H_0 + \sum_{k} \Psi^{\dag}_k\left( \begin{array}{ll}
a_k & b_k\\
b_k & a_k
\end{array} \right) \Psi_k,
\label{eq:eq3}
\end{equation}
where, $H_0=2NJ_1+2NJ_2-2NJ_3$ is the ground state energy and $\Psi_k^{\dag} = (b_{k}^{\dag},b_{-k})$,
where
\begin{eqnarray*}
a_k &=& J_1(\cos k_x -1) + J_3(\cos 2k_x-1) + J'_3\\
b_k &=& J'_3\cos k_y
\end{eqnarray*}

\begin{figure}[tbp]
\includegraphics[width=9cm, height=7cm]{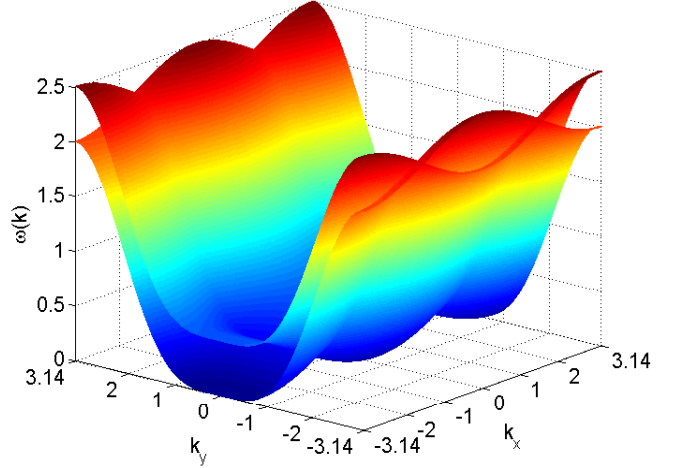}
\centering
\caption{(Color online) The spin wave dispersion relation as a function of $k_x$-$k_y$ in the commensurate phase $(0, \pi)$ CAF state for $2212^*$.
Parameters chosen are ($J_1, J_3, J'_3$) = $(-1, 0.25, 0.25)$.}\label{fig:fig5}
\end{figure}

Using the Bogliubov transformation, the linear spin-wave approximation, Eq. (\ref{eq:eq3}) can be diagonalized and shown in Fig. \ref{fig:fig5}. It is interesting to see the effect of $J_3$ on the spin wave excitations. For $J_3$ being AFM and close to $0.25J_1$, the spin wave dispersion along chain direction $(k_x,\pi)$ becomes quadratic at $k_x=0$. Otherwise, the dispersion is linear.

\subsection{Magnetic model for BaFe$_2$Se$_3$}

For the $123^*$ system, we can begin with the following general   Hamiltonian,
\begin{eqnarray*}
{\hat H} &=& J_{1b}\sum_{\langle i,j\rangle}\vec{S}_i\cdot\vec{S}_j + J_{1a} \sum_{\langle i,j\rangle}\vec{S}_i\cdot\vec{S}_j+ J'_{1a}\sum_{\langle i,j\rangle}\vec{S}_i\cdot\vec{S}_j
\notag\\
& + & J_2 \sum_{\langle\langle i,j\rangle\rangle}\vec{S}_i\cdot \vec{S}_j + J_3 \sum_{\langle\langle\langle i,j\rangle\rangle\rangle} \vec{S}_i\cdot \vec{S}_j + J'_{3} \sum_{\langle i,j\rangle}\vec{S}_i\cdot \vec{S}_j,
\end{eqnarray*}
where $J_{1b}$ along with $J_{1a},J_{1a}'$ and $J_3$ denote the intraladder vertical and horizontal NN and the third NN couplings, $J_2$ is the intraladder diagonal coupling and $J_3'$ is the interladder interaction as shown in Fig. \ref{fig:fig3}(c). These coupling parameters reflect the symmetry breaking of the BAF state.

We can treat the above model classically to obtain the exact ground state and phase diagram. We define the relative polarization angles ($Q_x,Q_x',Q_y,Q_y'$) along the different directions, with the primed variables going with the respective primed couplings. We can then write off the classical ground state energy as (for simplicity, we also take $S=1$)

\begin{eqnarray*}
E_c &=& 2 J'_{3}\cos Q_y'  + 2 J_{1b}\cos Q_y + 2 J_{1a}\cos Q_x \\
    &+& 2 J_{1a}'\cos Q_x' + 2 J_2\cos Q_y\cos Q_x'\\
    &+& 2 J_2\cos Q_y\cos Q_x + 4 J_3\cos(Q_x+Q_x')
\end{eqnarray*}

We can then obtain the ground states by simply minimizing the classical energy. With the BAF state  being the ground state, we have $ (Q_x,Q_x',Q_y, Q_y')=(0,\pi,0,\pi)$. Following the exchange coupling values measured for FeTe\cite{Lipscombe} and K$_{0.8}$Fe$_{1.6}$Se$_2$\cite{MWang}, we expect that
$J_{1a}\sim J_{1b}<0$, $J_3>0$, $J_3'>0$, $J_2>0$, and $J'_{1a}>0$. The strength of the couplings satisfies, $|J_{1a}|>J_2>J_3, J_3',J_{1a}'$, which stabilizes the BAF phase.

In the BAF state, we can obtain the spin wave excitations as done previously, which is given by
\begin{equation}
H = H_0 + \frac{1}{2}\sum_{k} \Psi^{\dag}_k \left( \begin{array}{ll}
A_k & B_k\\
B_k & A_k
\end{array} \right) \Psi_k,
\label{eq:eq5}
\end{equation}
where $H_0=NJ_{1a}+NJ_{1b}-NJ'_{1a}-2NJ_3-NJ'_3$ is the ground state energy and $\Psi_k^{\dag} = (b_{1k}^{\dag},b_{2k}^{\dag},b_{3k}^{\dag},b_{4k}^{\dag},b_{1,-k},b_{2,-k},b_{3,-k},b_{4,-k})$. $A_k$ and $B_k$ are four-by-four matrices, defined by:

\begin{subequations}
\begin{equation}
A_k =\left(
         \begin{array}{cccc}
           E_0 & J_{1a} & J_2 & J_{1b} \\
           .   & E_0 & J_{1b} & J_2 \\
           .   & .   & E_0 & J_{1a} \\
           .   & .   & .   & E_0 \\
         \end{array}
       \right)
\end{equation}
\begin{equation}
B_k = \left(
         \begin{array}{cccc}
           E'_0 & J'_{1a}e^{-ik_x} &  J_2e^{-ik_x} & J'_3e^{-ik_y} \\
           .    & E'_0 & J'_3e^{-ik_y} & J_2e^{ik_x} \\
           .    & .    & E'_0 & J'_{1a}e^{ik_x} \\
           .    & .    & .   & E'_0 \\
         \end{array}
       \right)
\end{equation}
\label{eq:eq6}
\end{subequations}
where $E_0 =-J_{1a} -J_{1b} + J'_{1a} + J'_3 + 2J_3$ and $E'_0=2J_3\cos k_x$. The lower triangle elements are suppressed because both matrices are hermitian.

\begin{figure}[tbp]
\includegraphics[width=9cm, height=7cm]{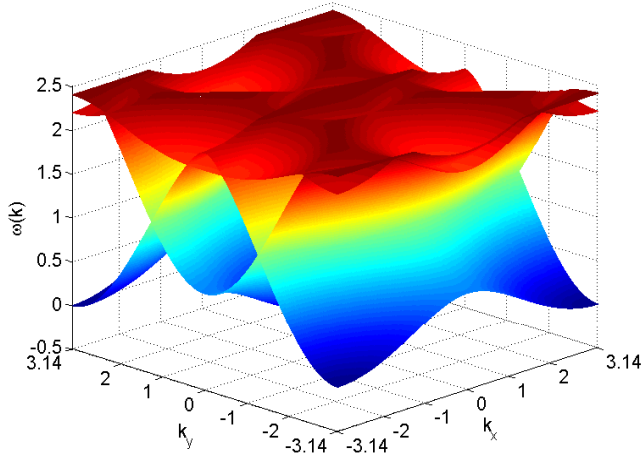}
\centering
\caption{(Color online) The spin wave dispersion relation of the lowest three branches (the other one is too high to be drawn in the same plot) as a function of $k_x$-$k_y$ in the BAF state for $123^*$. The chosen parameters are fixed as ($J_{1a}, J'_{1a}, J_{1b}, J_2, J_3, J'_3$) = (-1,0.1,-1,0.5,0.3,0.3).}\label{fig:fig6}
\end{figure}

By diagonalizing this Hamiltonian Eq. \ref{eq:eq5} for each $k$ in the Bosonic metric, we obtain the spin wave dispersion shown in Fig. \ref{fig:fig6} by taking $J_{1a}=J_{1b}=-1,  J_2=0.5, J_3=J_3'=0.3, J_{1a}'=0.1$. The spin wave has four branches which is very similar to the BAF state discussed for KFe$_2$Se$_2$\cite{Liwei}.

\section{Conclusion}

In this paper, we have performed the first-principles calculations for the electronic band structures and magnetic orders in these iron selenide systems featured with quasi-1D Fe chain (ladder) structures including $2212^*$ and $123^*$. However, the calculated FS topologies still exhibit two dimensional or even three dimensional features. For $2212^*$, we find that the ground state is a CAF ordered semiconductor with an energy gap of 0.64eV, in agreement with the experimental measurements. For $123^*$, the calculated results show that the ground state is a BAF ordered semiconductor with an energy gap of 0.24eV. These results suggest that that all magnetic structures in iron selenide compounds can be unified into an extended  $J_1$-$J_2$-$J_3$ model. We also calculate spin-wave excitations of the model which can be tested in future experiments on these two systems.

\section*{Acknowledgement}

We thank H. Ding, D. L. Feng, P. C. Dai, N. L. Wang, H. H. Wen, C. Fang and Uday Kiranfor  for useful discussion. W.L. gratefully acknowledges the financial support by Research Fund of Fudan University for the Excellent Ph.D. Candidates. The work was supported by the 973 Projects of China (2012CB821400) and NSFC-11190024.

\end{document}